\documentclass{article}

\usepackage{arxiv}

\usepackage[utf8]{inputenc} 
\usepackage[T1]{fontenc}    
\usepackage{hyperref}       
\usepackage{url}            
\usepackage{booktabs}       
\usepackage{amsfonts}       
\usepackage{nicefrac}       
\usepackage{microtype}      
\usepackage{lipsum}		
\usepackage{graphicx}
\usepackage{natbib}
\usepackage{doi}
\usepackage{appendix}
\usepackage{caption}

\title{How does course recommendation impact student outcomes? Examining directed self-placement with regression discontinuity analysis}

\date{April 9, 2025}	

\author{ \href{https://orcid.org/0000-0002-1977-9427}{\includegraphics[scale=0.06]{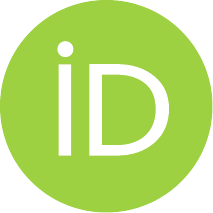}\hspace{1mm}Jason Godfrey}\\
	Strategic Data Fellow\\
	Harvard University\\
	Boston, MA 02138 \\
}

\renewcommand{\shorttitle}{\textit{arXiv} Template}

\hypersetup{
pdftitle={A template for the arxiv style},
pdfsubject={q-bio.NC, q-bio.QM},
pdfauthor={David S.~Hippocampus, Elias D.~Striatum},
pdfkeywords={First keyword, Second keyword, More},
}

\begin{document}
\maketitle

\begin{abstract}
	For many students, placement into developmental education becomes a self-fulfilling prophecy. Placing college students into developmental education significantly negatively impacts student attainment, student probability of passing, and college credits earned. To combat these negative effects, many universities are investigating alternative placement mechanisms. Could directed self-placement be an effective alternative mechanism? Do students who self-place suffer the same negative impacts from placement recommendations as their traditionally placed counterparts? This paper uses longitudinal data with causal inference methods to examine whether directed self-placement has similar negative impacts on student grades and pass rates as mandatory placement schema. We begin with an analysis of over 20,000 student placement records into one of two different placement tracks for first-year writing. Longitudinal and institutional data allow us to control for characteristic variables such as student race, family income, and sex. The results of our regression discontinuity design show that directed self-placement does not negatively impact student grades or pass rate. This may be an improvement for students who place at or near the threshold for developmental/remedial education; However, class, race, and gender-based statistical differences persist in the program at-large, demonstrating that placement technique plays only one part in building a more equitable program.
\end{abstract}

\keywords{Directed Self-Placement \and regression discontinuity \and writing program administration \and college placement \and remedial education \and education policy}

\section{Introduction}

First-year writing (FYW) is often a university's first opportunity to scaffold a new student's writing practices. According to the US Department of Education, eighty-five percent of students will take FYW before graduation. This makes it the most in-demand class in American post-secondary education by more than fifteen percentage points, and it is also the only course to see an increase in proportional enrollment over the last five decades. Currently, first-year writing is at an all-time high both in terms of absolute enrollment and proportional representation \citep{adelman_empirical_2004}.

But not all FYW classes are created equal. Many classes claim labels such as `honors' or `remedial.' These distinctions, observable at many universities across the United States and beyond, generate multiple possible paths through a first-year writing program. This trend towards creating separate courses depending on perceived student ability is growing. Every year US-based higher education enrolls an estimated 1.3 million students into developmental education \citep{pope_billions_2008}. As a proportion, this means that somewhere between one-third and two-fifths of all entering college students are placed into some form of developmental or remedial education, and this proportion has grown in recent decades \citep{greene_public_2003, snyder_digest_2016}. Bear in mind, this is the number of students who actually enroll in developmental education, and the number of students who are recommended such education in certain contexts may be as high as sixty percent \citep{bailey_challenge_2009}. In general terms, this high level of enrollment in developmental education indicates that universities deem a portion of their students as under-prepared to enroll in standard courses. The students in these courses fill a liminal space at the university setting: admitted but not fully accepted.

The implications of the `developmental' label echo into many disciplines and often follow students past their tenure in higher education. Existing literature and litigation use a gradient of labels for developmental education such as remedial, developmental, or transitional. These not-fully-synonymous labels often mean different things in different contexts; however, they all execute policies that divert students away from maximally efficient degree progress to take additional coursework—often for no or fewer credits. Current literature examining the causal impact of remedial education on student progress proposes that these courses yield mixed or poor results \citep{calcagno_impact_2008, bettinger_addressing_2009, scott-clayton_development_2014, boatman_does_2018}. Particularly, students placed into developmental education suffer lower probability of passing, fewer college credits earned, and lower attainment in college \citep{valentine_what_2017}. Beyond the difficulties of increased workload and more student debt, students placed into remedial education receive insignificant benefits to labor market outcomes \citep{martorell_help_2010}.

But what happens if students aren't ``placed'' into remediation? This study tests for similar outcome inequities in a context where students place themselves. In a system that uses directed self-placement (DSP), the student is the arbiter of their general education. A university will still provide resources for a student to gauge their probability of success in a course, create materials that advertise the rigor of different courses, and recommend a specific course based on individual preparation and academic background. However, ultimately, the student can register for whichever course seems best for them. We'll use regression discontinuity to examine the causal impact of directed self-placement. Our data comes from approximately 23,000 students who enrolled in a large public university from 2014-2019. This dataset allows us to explore students who are placed into developmental education and how their decision to follow or not follow the recommendation impacted their academic trajectory. Specifically, we'll examine how being placed into developmental education impacted final grades, and probability of failure.

The results of this study indicate that directed self-placement has no or insignificant causal impacts. Students placed into remedial education under this system are equally likely to succeed, enroll with similar frequency, and receive statistically similar grades. Consequently, directed self-placement allows for students wrongly assigned to remediation, and most at-risk for negative outcomes \citep{scott-clayton_development_2014}, to take standard courses, while students with lower levels of preparation who may benefit from remedial education can adhere to the university's recommendation. Although several qualitative and descriptive papers have posited such outcomes for directed self-placement, this study is the first to provide causal evidence for such claims.

\section{Literature review}

\subsection{Difficulties of developmental education}

Developmental or remedial education is a thoroughly studied aspect of higher education, and it is often found to have detrimental short- and long-term impact on student outcomes. One mounting challenge of developmental education in a writing context is its inherent characterization of students as deficient. Framing students as deficient has a long history \citep{valencia_evolution_2012} and is 'pervasive in higher education' \citep{heinbach_dismantling_2019}. In many ways bifurcating students into a standard educational track or a developmental track perpetuates already ingrained notions of a theoretical singular English that students must adopt to be considered a college-level thinker. Smit argues that `we need to find more suitable responses to diversity in the student body.' Smit goes on to assert that `describing ``disadvantage'' primarily in terms of poverty or socio-economic status gives an under-nuanced perspective. Employing a deficit mindset to frame student difficulties acts to perpetuate stereotypes, alienate students from higher education and disregards the role of higher education in the barriers to student success' \citep[p. 378]{smit_towards_2012}. These barriers to success have been observed in writing contexts across the academy, including in multilingual contexts \citep{rowan_transforming_2017, ruecker_intersection_2018}, first-year writing classrooms \citep{rowan_transforming_2017}, and library instruction \citep{heinbach_dismantling_2019}. Pedagogy-based solutions have been proposed and observed at a phenomenological level in at least as many contexts.

The study of how a deficit mindset to student placement negatively impacts student outcomes stretches into many corners of academia with many different terms and lines of argument. For \citet{martorell_help_2010}, the consequences of college remediation can be weighed and observed in tangible labor market outcomes. For \citet*{poe_reflection_2019}, the consequences are seen in first-year writing classrooms across the nation. For Hodara and Horn, the negative consequences are especially observable in a community college setting \citep{hodara_language_2012, horn_remedial_2009}. Although these researchers employ myriad methods of assessment, they converge on discontentedness with placement structures that define some students through supposed deficit as `developmental.'

Beyond promoting a deficit mindset, developmental education often doubly disadvantages the students that it is supposed to assist. This too has been studied in many contexts. To limit this literature review to only the most relevant research, we will focus solely on literature that investigates the impact of developmental education through an RD Design, which is the same method employed in this study. A recent meta-analysis of regression discontinuity studies in developmental education states `that placement into developmental education results in statistically significant and substantively sizable negative impacts' \citep[p. 826]{valentine_what_2017}. To make these findings even more relevant to this study, they also found evidence to `suggest that the negative effects of placement into developmental education are stronger for university students than for community college students and worse for students placed in reading or writing' (806). The educational context for this study: university-level writing.

\subsection{Directed self-placement as a potential solution}

Thus far, research on the impact of remedial education has been done in mandatory placement schema. There are always students who break institutional norms and enroll in courses other than those they are recommended to take. However, the rigidity of the placement procedures of the schools in previous studies are such that noncompliance is a variable that needs to be accounted for rather than an institutionally acknowledged procedure. Additionally, levels of noncompliance are quite low relative to noncompliance in a DSP system. For example, student non-compliance for college-level English was assessed to be at 17\% in `Shaping Policies Related to Developmental Education' \citep{moss_shaping_2006}. In some studies, non-compliance is not even addressed and may be low enough to be a non-issue \citep{duchini_is_2017}.

In the DSP model, instead of being placed into classes, students are informed about potential options and choose for themselves \citep{royer_directed_1998}. This may sound like a quotidian solution. But the stakes are high \citep{malenczyk_retention_2017, mutnick_basic_2011, owens_efficacy_2005}. Failure rates may increase if students put themselves in advanced courses. Additionally, less confident students may put themselves into lower classes even if they are qualified for the standard educational track, incurring greater tuition fees and delaying major-focused coursework. Also, there is more research and precedent behind standard placement models. DSP is young. It has less research. There are fewer examples to template a program from. Simple institutional inertia is enough to limit the adoption of DSP unless it shows demonstrable, replicable benefits. When placement impacts thousands of students every year, and when those changes can mean higher failure rates, increased student debt, or decreased student experience, every change matters \citep{royer_directed_1998}.

DSP began in 1998 at the GVSU writing program \citep{royer_directed_1998}. Since then, it has mostly been implemented in other writing programs. So, most of its research focuses on writing programs. Nevertheless, DSP is also used in STEM education and in international contexts \citep{mendez_motivacion_2016, bracco_college-ready_2019, burn_national_2018, mejia_what_2019, burdman_degrees_2015, felder_informed_2007, kosiewicz_giving_2020, brown_technology_2010}. Research beyond a US-based writing context is developing; one study found only 2\% of mathematics departments use DSP \citep{burn_national_2018}, and research mentioning DSP in an international context doesn't focus on the placement method itself \citep{mendez_motivacion_2016}. However, DSP offers distinct affordances from mandatory placement schema. Ideally, DSP allows students to become agents in their educational trajectory \citep{moos_directed_2019}. An individual student also knows their own educational history. When students can choose their classes, they divert power from the institution, and get a better say in their own education. Unfortunately, a better say doesn't always mean better outcomes. DSP doesn't guarantee that students will place optimally. Rather, it empowers students to author their own educational experience.

How these conditions are met is an area of on-going research and debate. California and Florida are two states where legislation partially dictates placement procedure. Florida's Senate Bill 1720 states that remedial education can't be mandatory for most students. And California's Executive Order 1110 cut non-credit-bearing remedial education. These legislative ecologies demand self-placement. Whether this self-placement is "directed" has arguments on both sides. If these relatively new placement ecologies qualify as directed self-placement, data from these states could be used to validate DSP at state-wide levels. If not, that data is not a candidate for DSP research. This is an important debate since DSP research so far has been conscientiously phenomenological \citep{balay_placing_2012, nastal-dema_reenvisioning_2014, toth_directed_2014}. This means that much current research focuses on student experiences. This article will be among the first to empirically examine DSP for causal impact on student outcomes.

\section{Methods}

\subsection{Institutional context}

This paper uses data from a large, public, 4-year+ university in the Midwest. The sample used in this study is from the university's largest college, which enrolls approximately 4,500-5,000 students in first-year writing annually. It should be noted that the average high school GPA for this sample is above a 3.8 and the historic graduation rate from the university is over 90\%. The general homogeneity at the institutional level on these and other student characteristics will contribute to low levels of dispersion across many characteristic and outcome variables. The large sample size provides enough statistical power to compensate for low dispersion. Perhaps of greater concern may be the temptation to be dismissive of small significant changes as unsubstantial. Because of the relatively narrow distributions, we may consider some differences substantial that in other contexts may not be considered substantial.

Most of the data used in this study comes from an institution-level repository developed specifically for helping researchers answer questions about learning analytics. This data was supplemented by a dataset that contained responses to the DSP questionnaire (the internally developed tool used to recommend students into traditional or remedial education) and another dataset that contained a list of courses that each student had taken that fulfill a writing requirement and another dataset that would assist in merging the previous datasets. Once all datasets were merged, complete records remained for 21,729 students out of the initial dataset of 22,648.

This research was done on student data from the years 2014-2019. This is because during these five years the DSP questionnaire was administered without alterations. This ensured that comparisons across academic semesters were examining identical sets of questions, and that any endogenous changes that might impact student placement were accounted for. Having a recent 5-year sample has advantages. The data is recent and is sure to reflect current placement trends. Additionally, current program administrators could field questions about information missing in data and documentation. A recent sample also has disadvantages. Several outcomes that are common in similar studies are not available here. For example, time-to-degree and grades in advanced courses are often considered as outcome variables. Due to recency, those incorporating these variables would sacrifice necessary statistical power. To illustrate, of the 21,795 students in the final dataset, advanced writing grades were only available for 2,574 students. However, since student maturation between initial outcome variables and delayed outcome variables is itself a threat to validity, such omissions may not be so harmful.

\begin{figure}
  \centering
  \includegraphics[scale=0.75]{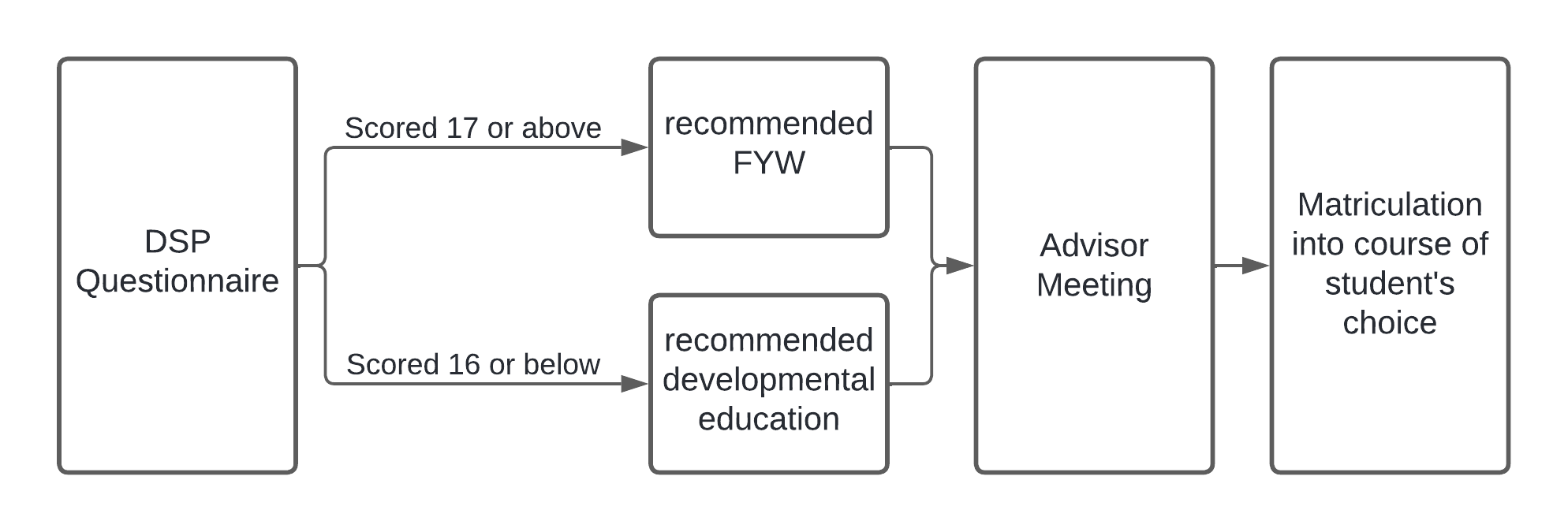}
  \caption{The placement process at the studied university.}
  \label{fig:placement-process}
\end{figure}

To be placed into first-year classes students at this university go through an orientation where they complete multiple in-house tests and speak with an advisor. Once the advisor meeting is finished, the student may enroll on their own. The in-house test for the writing program is a directed-self placement questionnaire, which asks students about previous writing tasks and their comfort with writing in English at a college-level. Students know that their performance on pre-enrollment tests and questionnaires influences placement recommendations, but they are not informed about grading criteria or any of the specific thresholds for placement. We want to take special note that the questionnaire is not a skill check; the questions aren't about grammar or familiarity with standard research or citation protocols. Instead, the questionnaire focuses on student experiences, practices, and self-beliefs—as is standard practice \citep{toth_directed_2014}. The students who score at 17 or above on this questionnaire are recommended to take standard first-year writing. The students who score 16 or below are recommended to take one of two pass/fail, for-credit writing classes that do not fulfill the first-year writing requirement. Students enrolled in the pass/fail courses will have to take FYW after completing the pass/fail course. This process is illustrated in figure \ref{fig:placement-process}.

There are two developmental courses. There is one developmental writing course for students who are comfortable with English but would like additional support for college-level writing, and there is a separate course for students who would like additional support with English. There are 22 different first-year writing courses that are documented in the data. They all fulfill the same prerequisite, but they are hosted by different majors such as classics, biology, or art and design. Despite the variety of course offerings, most students who enroll in first-year writing elect to take a general writing course that will be referred to as `ENG 101.' Although students receive a recommendation and meet with an advisor to discuss placement options, they are free to enroll in whichever course they choose. In short, a student who is recommended to take developmental education can opt out with minimal institutional friction, and, as the data will show, many do just that.

\section{Regression discontinuity design}

This study uses regression discontinuity (RD) design to search for causal impact of placement into developmental education. This RD compares students who scored just above and just below the threshold for FYW recommendation. The first stage of the analysis was to seek evidence that the assumptions necessary for valid RD design were met.  These assumptions are as follows: 1. Density and continuity hold across assignment threshold; 2. Assignment has been followed; 3. No unobserved or coincidental factors impact placement around the threshold examined \citep{trochim_research_1984, cook_empirical_2008, mccall_regression_2012}. If all these assumptions are met, RD design can produce unbiased estimates of the average treatment effect for compliers at/near the cutoff threshold.

The first assumption of RD analysis requires that differences in student characteristic variables are as good as random across the threshold. Or, RD assumes ceteris paribus on characteristic variables and that students differ only in placement \citep{calonico_robust_2014}. For this study, I'll look for evidence of continuity across assignment threshold. Figure \ref{tbl:placement} shows students on either side of the threshold. The table shows descriptive statistics for multiple bandwidths to check for robustness; however, the final  +/- 3 bandwidth selected for this study is justified in the later section ``Model choice and calculating bandwidth.'' Because RD focuses examination on students within proximity to the threshold, results cannot be applied to students who score extremely on either end of the cutoff.

The summary statistics in table \ref{tbl:placement} show some differences in students on either side of the threshold. Students placed into FYW are from families whose estimated gross income is a few thousand higher than students just below the threshold, for example. However, for the majority of observable characteristics students are relatively similar across the cutoff. This high level of cohesion across variables evidences sound internal validity for an RD design as far as threshold variation is concerned. Knowing that students are relatively homogenous across summary statistics through the threshold will allow the search for significant difference in outcome variables: FYW course grade and pass rate.

Another way to search for possible discontinuities in assignment across the threshold is through graphical interpretation. Figure \ref{fig:dsp_dist} below shows the distribution of DSP questionnaire scores by frequency.

\begin{figure}
  \centering
  \includegraphics[scale=0.50]{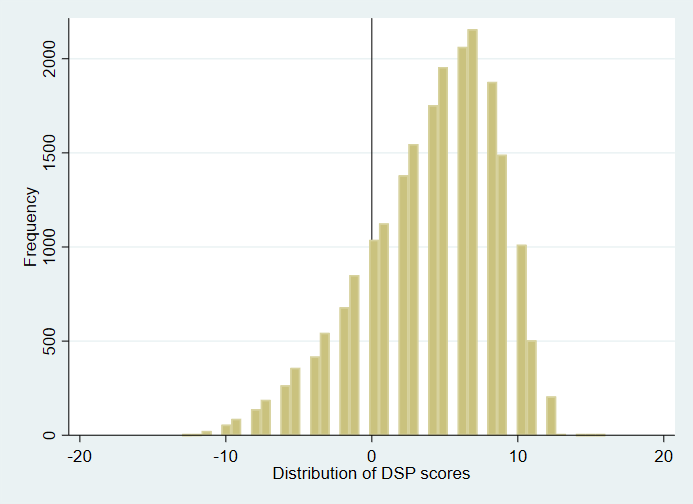}
  \caption{This distribution of student scores for the DSP questionnaire.}
  \label{fig:dsp_dist}
\end{figure}

Any discrepancy across the threshold could evidence participants manipulating their assignment to gain a desired outcome. In this case, we would probably expect to see a sudden jump in scores across the 17-score threshold. That would indicate that students are unnaturally inflating their scores just across the threshold to avoid a placement recommendation into remedial education. The kurtosis-skewness test for this dataset is p$>$z = 0.0001. The above histogram shows the distribution around the threshold. The kurtosis-skewness test shows that the data are normally distributed, and the histogram provides a visual indication of that fact. Another robust check of this assumption is the McCrary density test, which was designed specifically to test for continuity of the running variable density function \citep{mccrary_manipulation_2008}. This test is available in the figure \ref{fig:dsp_mcrary} below. The results of this test show no evidence of bunching of scores around the cut-off for placement (theta = 0.0026, s.e. = 0.042, z = 0.032); this data set passes the density assumption required for RD analysis.

\begin{figure}
  \centering
  \includegraphics[scale=0.50]{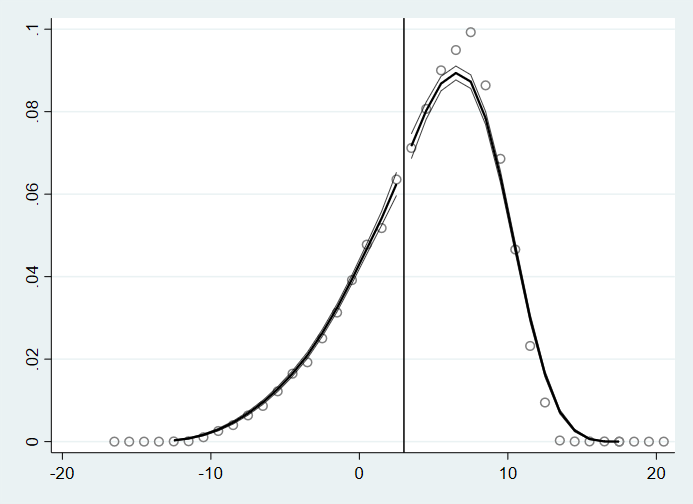}
  \caption{The placement process at the studied university.}
  \label{fig:dsp_mcrary}
\end{figure}

To discuss assumption 2, that assignment has been followed, it will be helpful to look at how directed self-placement impacts enrollment patterns. Table \ref{tbl:enrollment_proportion} enumerates how students matriculated according to their placement recommendation. Because this study takes place in the context of a directed self-placement structure, high levels of non-compliance with recommendations are not a threat to internal validity.

\begin{table}[!ht]
    \centering
    \caption{Student matriculation by course recommendation}
    \begin{tabular}{llll}
    \hline
        DE Recommended & All & Took DE & Didn't take DE \\ \hline
        0 & 0.834 & 0.037 & 0.963 \\ 
        1 & 0.166 & 0.704 & 0.296 \\ \hline
    \end{tabular}
\label{tbl:enrollment_proportion}
\end{table}

This table shows a couple of interesting trends. First, 16\% of students were recommended developmental education, but only 6.3\% of students enrolled. In numerical terms, this equates to 2,234 students defying their placement recommendation and only 1,377 students complying. Another interesting trend is that only 95.8\% of students with 3 points above the threshold did not take developmental education. As previously stated, this non-compliance provides context for how liberally students choose to self-place above and below their assigned course level, but it does not impact study design. In other words, recommendation to remedial education does not equate to intent-to-treat (ITT); that's the point of DSP. Because this study seeks to examine the impact of labelling a student as in need of “remediation” rather than the impact of administered remediation itself, issues of compliance and non-compliance, although necessary context, are not necessarily relevant to the RD design itself. This table confirms that 100\% of students above the cutoff received no remedial recommendation, and 100\% of students below the cutoff received a remedial recommendation; consequently, this study adopts a sharp regression discontinuity design.

Assumption 3 cannot be tested directly; exogeneity is inherently unprovable. Evidence for it can be searched for, however. The manner that this assumption is tested depends on the results of the RD analysis. More on assumption 3 and exogeneity will therefore be discussed in the subsequent section.

\subsection{Model choice and calculating bandwidths}

To determine model choice for this RD study linear, cubic, and quadratic models were considered. The relationship between DSP questionnaire score and eventual FYW grade were found to follow a linear pattern that is significantly correlated with DSP score. Using standard notation, the model for the first test in this study is as follows:
$$Y_i = \beta_0 + \beta_1(DSP_i) + \beta_2(Z) + \beta_3(DSP_i)(Z) +\epsilon_i$$

Where Y represents the outcome grade for any student $i$. The variable $Z$ represents a binary variable of students scored below the placement threshold (17). The variable $DSP_i$ is the DSP questionnaire score. Finally, $\epsilon_i$ is the residual term.

The relationship between DSP questionnaire score and probability of passing FYW on first attempt also followed a linear pattern, using standard notation, the model for the second test in this study is identical to the previous notation, except $Y_i$ indicates probability of passing as a dichotomous variable where $0$ = failure, incomplete, or withdrawal and $1$ = any passing grade.

The bandwidth was calculated according to nearest neighbor-based variance estimators exposited in \citet{imbens_optimal_2012} and discussed at length in \citet{calonico_robust_2014}.  Using this method, the bandwidth estimation was set to $\pm$ 3.683. This is the source for why previous tables in the methods section have examined scores within 3.683 points of the threshold. This bandwidth was tested for sensitive results when control variables were eliminated. We also tested by increasing the bandwidth by up to ±6 points. In both cases, there was no significant effect on the estimates.

\section{Results}

This section presents results of the RD design based on two separate outcome variables: grade in first-year writing and probability of passing first-year writing on the first attempt. For each of these outcome variables, we estimate discontinuities for the following seven characteristic variables: sex, age, race, estimated gross family income, ACT score (\%ile English, \%ile total), and in-house math placement score.

Figure \ref{fig:characteristic} in the appendix contains the output for selected characteristic variables across the placement threshold. If the effect of being labelled as a student in need of remedial education were substantial, there would be a discontinuity in the regression across the threshold. In these particular graphical representations, the confidence intervals are represented by grey lines extending vertically from each point along the regression line. These graphs show that along the threshold, none of the selected characteristic variables show discontinuities beyond the limits of a .05 confidence interval. Gross family income approaches discontinuity, but its estimations are still within the limit. Tabular output including standard error, $z-score$, and $P > | z |$ appear below in table \ref{tbl:rdrace}.

Noticeably missing from the characteristic variables examined in the appendix is any breakdown along the lines of racial representation. These will be presented in \ref{tbl:rdrace} below. The reason for emphasizing tabular format when one of the affordances of RD design is easily interpretable graphical results stems from the highly variable sample size among races. The studied university is predominantly white, and some racial minorities have disproportionately low representation, and these inequities are more clear when the sample sizes can be quickly juxtaposed.

\begin{table}[!ht]
    \centering
    \caption{RD Estimates with $n$ per self-selected racial categories}
    \begin{tabular}{lllllll}
    \hline
        ~ & Asian & Black & Hawaiian & Hispanic & Native American & White \\ \hline
        RD Estimate & 0.0458 & -0.0276 & 0.00383 & -0.0134 & 0.00186 & -0.0437 \\ 
        ~ & (0.0713) & (0.0224) & (0.00456) & (0.0213) & (0.00782) & (0.0399) \\ 
        Observations & 4,428 & 1,346 & 76 & 1,335 & 234 & 15,420 \\ \hline
    \end{tabular}
    \caption*{\footnotesize Notes: Standard errors in parentheses, * p $<$ 0.05, ** p $<$ 0.01, *** p $<$ 0.001}
    \label{tbl:rdrace}
\end{table}

As this table demonstrates, no discontinuities near the cutoff can be observed from racial characteristics. None of these findings are significant. For most of these models, sample size is sufficient that any major differences would have been detected. In some cases, such as for Native Americans and for Hawaiian/Pacific Islanders, lack of statistical power prohibits interpretation.

The table below details estimations for the outcome variables of this study. Namely, they show (1) student grade and (2) percent chance of passing FYW on the first attempt as they move across placement category thresholds. The coefficients and p-values of the accompanying underlying regressions are reported in table \ref{tbl:multi_regression}; no significant discontinuities are estimated.

\begin{table}[!ht]
    \centering
    \caption{Coefficients and p-values of characteristic variables}
    \begin{tabular}{lllll}
    \hline
        ~ & RD Estimate & Std. Err. & Z-Score & P $>$ z \\ \hline
        Female & .00112 & .0418 & 0.0269 & 0.979 \\ 
        Age & -.01288 & .061 & -0.2111 & 0.833 \\ 
        Family Income & -14.561 & 7.1174 & -2.0458 & 0.051 \\ 
        HS GPA & -.08402 & .09434 & -0.8906 & 0.373 \\ 
        ACT Score & -.28384 & .31833 & -0.8917 & 0.373 \\ 
        ACT English \%ile & -1.3972 & 3.0926 & -0.4518 & 0.651 \\ 
        Math Placement & -.26995 & .46997 & -0.5744 & 0.566 \\ 
        Pass FYW & .00199 & .00798 & 0.2487 & 0.804 \\ 
        FYW grade & -.04136 & .53029 & -0.0780 & 0.938 \\ \hline
    \end{tabular}
    \caption*{\footnotesize Notes: Standard errors in parentheses, * p $<$ 0.05, ** p $<$ 0.01, *** p $<$ 0.001}
    \label{tbl:rdrOutput}
\end{table}

In both outcome variables, there is no significant estimated discontinuity. Regarding educational outcome by DSP performance, using OLS regression there is a strong positive correlation between DSP score and grade in FYW (p $<$ 0.000). For probability of passing FYW, using logistic regression there is a strong positive correlation between DSP score and probability of passing FYW on first attempt (p = 0.006). However, for both of these outcomes there is no detectable discontinuity across the placement threshold.

\section{Discussion and limitations}

If assumptions were properly satisfied, correct models were applied, and causal inference can be applied, then this study provides evidence that recommending developmental education does not negatively impact predefined student outcomes for students near the placement threshold, given that students may autonomously self-place.

\begin{figure}[ht]
    \centering
    \begin{minipage}{0.48\textwidth}
        \centering
        \includegraphics[width=\linewidth]{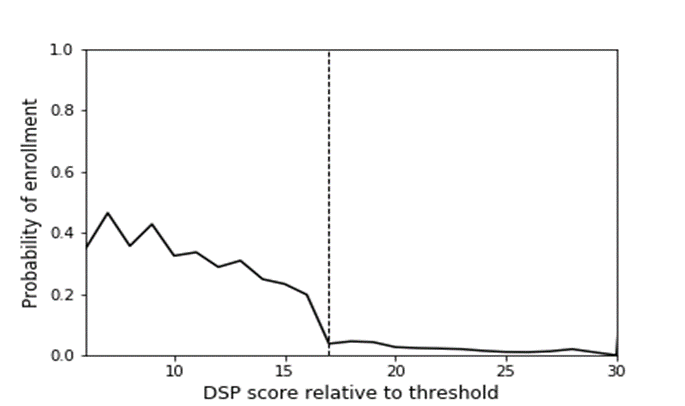}
        \caption{Student Probability of following placement at studied institution}
        \label{fig:dsp-follow}
    \end{minipage}
    \hfill
    \begin{minipage}{0.48\textwidth}
        \centering
        \includegraphics[width=\linewidth]{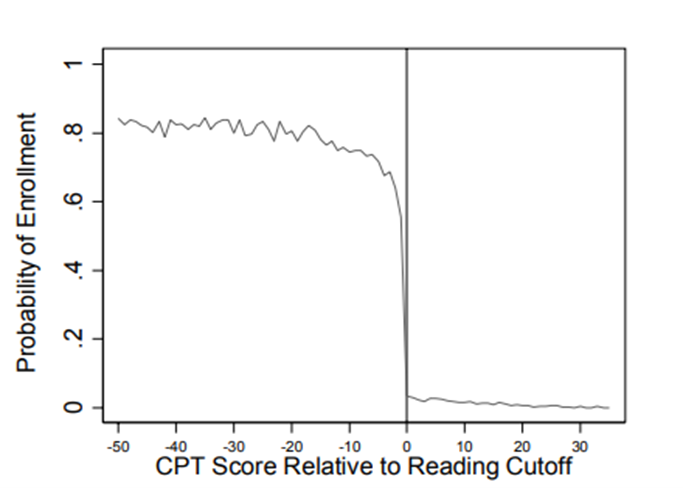}
        \caption{Student Probability of following placement in \citet{calcagno_impact_2008}}
        \label{fig:nondsp-follow}
    \end{minipage}
\end{figure}

Students placed in developmental education in a DSP system have the freedom to simply re-place themselves into whatever course they choose; this may be a contributor to the lack of negative discontinuous results found in this test. \citet{scott-clayton_development_2014} posit that the negative effects are most profound “for a subgroup we identify as potentially misassigned to remediation.” Otherwise, “remedial assignment does little to develop students' skills. But we also find little evidence that it discourages initial enrollment or persistence”. If this is true, then a student's ability to merely ignore institutional placement seems a possible solution. And students certainly do ignore placement when given the chance. Below are two figures. Figure \ref{fig:dsp-follow} shows student probability of enrolling in developmental education by DSP score in the program under study. Figure \ref{fig:nondsp-follow} is from another RD study in a program with mandatory placement \citep{calcagno_impact_2008}. These figures are placed side-by-side for easy comparison.


The trend in following placement recommendations is relatively similar across studies. Students with the lowest scores are the most likely to follow placement recommendation, and the probability drops as students approach the threshold. Above the threshold, probability of enrolling in developmental education drops precipitously and remains essentially 0. The difference in these two figures is that student probability of enrolling in developmental education if the student is recommended developmental education is nearly always around 40 percentage points lower at the studied institution than at the institutions examined in \citet{calcagno_impact_2008}. Similar probability charts to Boatman's can be found in other studies \citep{martorell_help_2010, boatman_evaluating_2012, hodara_language_2012, scott-clayton_development_2014}. DSP represents a substantial change in program administration that is plainly observable and seems a possible candidate for the changes in program outcomes.

\subsection{Neutral impact of placement assignment != neutral program Outcomes}

Although this paper provides causal estimates which evidence a neutral impact of placement assignment for students near the cutoff, that does not mean that the program under study produces neutral program outcomes. This next section examines program outcomes for all students, and not just students at or near the placement threshold. By zooming out to show outcomes across the entire student population, we hope to demonstrates how even empirically neutral programmatic placement can exist in a program with significantly unequal placement and learning outcomes for students based on income, race, and sex. 

Table \ref{tbl:multi_regression} outlines a multiple regression model using all the same characteristic variables from the RD design. The outcome variable is grade, also the same as with the RD design. This table considers students from the entire spectrum of test takers, not just students within proximity of the placement threshold.

\begin{table}[!ht]
    \centering
    \caption{Results of multiple regression model of FYW grade}
    \begin{tabular}{llll}
    \hline
        ~ & (1) & (2) & (3) \\ \hline
        DSP Test & 0.143*** & 0.143*** & 0.119*** \\ 
        ~ & (0.0115) & (0.0118) & (0.0128) \\ 
        Native English Speaker & 0.386 & 0.286 & 0.413 \\ 
        ~ & (0.299) & (0.307) & (0.715) \\ 
        ACT \%ile & ~ & ~ & 0.294*** \\ 
        ~ & ~ & ~ & (0.0209) \\ 
        HS GPA & ~ & 0.0692 & 0.0378 \\ 
        ~ & ~ & (0.0475) & (0.0435) \\ 
        Family Income & ~ & 0.0719* & 0.0295 \\ 
        ~ & ~ & (0.000436) & (0.000486) \\ 
        Asian & 0.376 & 0.403 & -0.307 \\ 
        ~ & (0.244) & (0.246) & (0.262) \\ 
        Black & -1.234*** & -1.163*** & -0.227 \\ 
        ~ & (0.311) & (0.313) & (0.364) \\ 
        Hawaiian & -0.673 & -0.996 & -0.593 \\ 
        ~ & (1.696) & (1.822) & (1.606) \\ 
        Hispanic & -0.666** & -0.631** & -0.367 \\ 
        ~ & (0.305) & (0.309) & (0.339) \\ 
        Native American & -0.400 & -0.356 & -0.655 \\ 
        ~ & (0.966) & (0.964) & (1.197) \\ 
        Not Indicated & 0.277 & 0.295 & -0.317 \\ 
        ~ & (0.287) & (0.296) & (0.335) \\ 
        White & 0.0637 & 0.0748 & -0.0938 \\ 
        ~ & (0.211) & (0.214) & (0.218) \\ 
        Constant & 86.67*** & 86.39*** & 78.18*** \\ 
        ~ & (0.474) & (0.501) & (1.251) \\ 
        ~ & ~ & ~ & ~ \\ \hline
        Observations & 20,160 & 19,649 & 14,993 \\ 
        R-squared & 0.015 & 0.015 & 0.036 \\ \hline
    \end{tabular}
    \caption*{\footnotesize Notes: For the results of the multiple regression. Standard Errors in Parenthesis. The omitted category is 'two or more ethnicities.' The dependent variable is grade., * p $<$ 0.05, ** p $<$ 0.01, *** p $<$ 0.001}
    \label{tbl:multi_regression}
\end{table}

The first consideration with this table is that on average, if the students scored 0 on the DSP questionnaire, they would still be expected to earn somewhere between a 78.18-86.67, depending on which characteristic variables you include. None of these scores are within failing range. Indeed, the average grade for this course is quite high ($~$90) and narrowly distributed. Nevertheless, the sample is large enough to provide statistical power and the grades follow a normal, if highly skewed, distribution. This is to say, that an argument against the substantive impact of this analysis could be made, since nearly 99\% of the students pass the course, and the average grade those students receive is an A-. However, in an environment where GPA is a major consideration for internal program applications and for graduate school, the difference between an 89 and a 93 may feel more substantial to a student than it appears in a multiple regression table. Further, in an ideal educational setting the differences between these grades would not be significantly predicted by a student's race, income, or sex. In the least inclusive regression, students who identified as black are associated with a grade that is 1.23 percentage points lower than their peers; this finding is significant at p$<$0.01. For students who identified as Hispanic the grade drop is .67 percentage points; this finding is significant at p$<$0.05. When estimated family income is included, each thousand dollar increase in family income is associated with .00072 point increase. In an extreme case of \$200,000 disparity, this would account for 14.4 percentage points of difference on average; this finding is significant at p$<$0.10. The final column of the regression table includes ACT percentile, which is significant at p$<$0.01. This column also omits 26\% of the sample; thus, its results should be interpreted with skepticism.

In all, this table provides evidence that even a program with a placement procedure with neutral impact can still produce unequal outcomes. Investigating the depths of these inequalities and proposing methods to mitigate them is beyond the scope of this study; however, we would be remiss not to pause on this area where additional research into diversity, equity, and inclusion could help further mitigate negative outcomes.

\subsection{Limitations and areas for future researchers}

Further research is needed in multiple areas. First, this study does not directly compare mandatory and directed self-placement systems. While \citet{kosiewicz_giving_2020} do have a study that directly compares these two placement systems in a mathematics education environment, it is the sole article capable of making empirically justifiable causal claims to do so. Second, this study focuses on a single institution with a high proportion of students who complete their FYW courses. While that may be good for many reasons, it complicates arguments about the substantive nature of many results. Third, this study was conducted with data that is too new for mid- and long-term outcomes to be included. Notable omissions include time to enrollment, time to graduation, and total credits earned. The inclusion of mid- and long-term outcome variables would increase the interpretative power of this study. It is possible that the positive or negative effects of developmental education aren't immediately apparent. If so, this study would fail to detect those effects.

\section{Conclusion}

Previous studies have shown DSP as a powerful set of tools that is highly flexible to local circumstances \citep{gere_local_2013}, should be assessed locally \citep{gere_assessing_2010}, and empowers students to make informed decisions \citep{toth_directed_2014}. This study asked a simple but consequential question: what happens when the power to decide rests with the student rather than the institution? Using a regression discontinuity design at a large public university, we found that placement into developmental education under a directed self-placement (DSP) system produced no significant discontinuities in course grades or pass rates. In other words, when students were free to override recommendations, being labeled “remedial” carried no detectable penalty near the placement threshold.

This finding matters. Prior studies of mandatory placement consistently show that remediation can depress student outcomes, particularly for those likely misassigned to remediation \citep{calcagno_impact_2008, bettinger_addressing_2009, scott-clayton_development_2014, valentine_what_2017}. By contrast, DSP appears to blunt that harm: students most at risk of negative effects can opt into standard courses, while those who prefer added support can take it. What had been theorized in qualitative accounts of DSP \citep{royer_directed_1998, toth_directed_2014, gere_assessing_2010} is here supported with causal evidence.

At the same time, a neutral causal estimate does not imply a neutral program. Our broader analyses show persistent inequities in outcomes by income and race, echoing long-standing concerns about deficit framings and systemic barriers in writing instruction \citep{valencia_evolution_2012, heinbach_dismantling_2019}. The lesson is twofold: placement policies like DSP can mitigate some of the structural disadvantages built into traditional remediation, but eliminating inequities requires deeper changes to curriculum, pedagogy, and institutional support.

Taken together, this study offers a methodological contribution bringing RD design to the study of DSP in first-year writing \citep[cf.][]{haswell_beyond_2001} and a practical one: institutions weighing alternatives to mandatory remediation now have preliminary causal evidence that DSP is a viable, equity-forward foundation for student choice.

To writing program administers, this study provides initial causally inferential evidence that DSP is a viable, equity-forward foundation for student choice. To administrators of non-writing programs, 

This study uses RD design to provide a causal estimate for the impact of recommendation for placement into developmental education on quantitative student outcomes. While such studies have been conducted previously, this study innovates on previous research by applying the popular methodology to a university that uses directed self-placement. Previous studies into the impact of developmental education have found that developmental education can have significant and substantial negative effects on students. This study failed to find such effects. Juxtaposing the efficacy of different placement methods is beyond the scope of this paper; however, the lack of discontinuously negative outcomes for students placed into developmental education with DSP provides preliminary causal evidence for the efficacy of such self-placement programs.

Additionally, it innovates on previous research studying outcomes in a first-year writing context by using an empirical design that is replicable, aggregable, and data-driven \citep{haswell_beyond_2001}. And, if we may add yet another qualification to Haswell's heuristic for empirical research in writing, this study has been specifically designed to measure the causal impact of a specific practice highly particular to writing-centric fields (DSP). Previous studies have shown DSP as a powerful set of tools that is highly flexible to local circumstances \citep{gere_local_2013}, should be assessed locally \citep{gere_assessing_2010}, and empowers students to make informed decisions \citep{toth_directed_2014}. This study contributes to this body of scholarship by empirically examining the causal impact of placement within such a system, a methodological contribution that has been a barrier to non-writing programs considering adoption \citep{burn_national_2018}.

\bibliographystyle{unsrtnat}
\bibliography{references}  






\section*{Declarations}

The authors of this paper declare that they have no conflict of interest.

\newpage
\begin{appendices}
\section{}\label{secA1}




\begin{table}[!ht]
    \centering
    \caption{Characteristic and outcome variables by RD bandwidth}
    \resizebox{\columnwidth}{!}{
    \begin{tabular}{llllll}
    \hline
        ~ & All & Below Threshold & Above Threshold & Difference & Effect Size \\ \hline
        \textbf{Panel A: All Students} & ~ & ~ & ~ & ~ & ~ \\ 
        Female & 0.568 & 0.510 & 0.580 & 0.070 & 0.141*** \\ 
        Age & 18.335 & 18.458 & 18.311 & -0.147 & 0.198*** \\ 
        Asian Reporting & 0.408 & 0.559 & 0.377 & -0.181 & 0.226*** \\ 
        Black Reporting & 0.062 & 0.089 & 0.057 & -0.032 & 0.133*** \\ 
        Hawaiian/Pacific Islander Reporting & 0.003 & 0.003 & 0.004 & 0.001 & 0.009 \\ 
        Hispanic Reporting & 0.061 & 0.068 & 0.060 & -0.008 & 0.033+ \\ 
        Native American Reporting & 0.011 & 0.010 & 0.011 & 0.001 & 0.006 \\
        White Reporting & 0.710 & 0.599 & 0.732 & 0.133 & 0.295*** \\ 
        Estimated Gross Family Income & 113.020 & 103.324 & 114.966 & 11.641 & 0.136*** \\ 
        High School GPA & 3.534 & 3.354 & 3.570 & 0.216 & 0.213*** \\ 
        Max ACT Score & 30.206 & 29.238 & 30.387 & 1.148 & 0.381*** \\ 
        Max English \%ile & 67.274 & 57.275 & 69.136 & 11.861 & 0.387*** \\ 
        Max In-House Math Placement & 17.564 & 17.370 & 17.602 & 0.233 & 0.044* \\ 
        Pass FYW on first attempt & 0.989 & 0.986 & 0.990 & 0.004 & 0.042* \\ 
        Grade in FYW & 90.120 & 89.050 & 90.283 & 1.233 & 0.219*** \\ 
        Native English Speaker & 1.961 & 1.916 & 1.970 & 0.054 & 0.281*** \\ 
        N & 21729 & 3611 & 18118 & ~ & ~ \\ \hline
        \textbf{Panel B: Bandwidth = 2} & ~ & ~ & ~ & ~ & ~ \\ 
        Female & 0.531 & 0.521 & 0.537 & 0.016 & 0.032 \\ 
        Age & 18.361 & 18.385 & 18.343 & -0.042 & 0.059+ \\ 
        Asian Reporting & 0.463 & 0.486 & 0.448 & -0.038 & 0.046 \\ 
        Black Reporting & 0.067 & 0.078 & 0.060 & -0.018 & 0.070* \\ 
        Hawaiian/Pacific Islander Reporting & 0.004 & 0.003 & 0.005 & 0.001 & 0.021 \\
        Hispanic Reporting & 0.066 & 0.070 & 0.064 & -0.006 & 0.024 \\ 
        Native American Reporting & 0.009 & 0.009 & 0.008 & -0.001 & 0.009 \\ 
        White Reporting & 0.662 & 0.640 & 0.677 & 0.037 & 0.078* \\ 
        Estimated Gross Family Income & 106.982 & 107.798 & 106.407 & -1.391 & 0.016 \\
        High School GPA & 3.477 & 3.432 & 3.508 & 0.076 & 0.070* \\ 
        Max ACT Score & 29.812 & 29.661 & 29.917 & 0.256 & 0.080* \\ 
        Max English \%ile & 62.117 & 60.689 & 63.118 & 2.429 & 0.076+ \\ 
        Max In-House Math Placement & 17.464 & 17.437 & 17.483 & 0.046 & 0.008 \\
        Pass FYW on first attempt & 0.992 & 0.991 & 0.993 & 0.002 & 0.019 \\ 
        Grade in FYW & 89.606 & 89.568 & 89.628 & 0.061 & 0.011 \\ 
        Native English Speaker & 1.948 & 1.937 & 1.956 & 0.018 & 0.083* \\ 
        N & 3694 & 1531 & 2163 & ~ & ~ \\ \hline
        \textbf{Panel C: Bandwidth = 3} & ~ & ~ & ~ & ~ & ~ \\ 
        Female & 0.532 & 0.519 & 0.541 & 0.022 & 0.044 \\ 
        Age & 18.356 & 18.391 & 18.336 & -0.056 & 0.078** \\ 
        Asian Reporting & 0.469 & 0.509 & 0.446 & -0.063 & 0.074** \\ 
        Black Reporting & 0.070 & 0.081 & 0.064 & -0.017 & 0.066* \\ 
        Hawaiian/Pacific Islander Reporting & 0.004 & 0.003 & 0.004 & 0.001 & 0.022 \\ 
        Hispanic Reporting & 0.066 & 0.066 & 0.066 & 0.001 & 0.003 \\ 
        Native American Reporting & 0.012 & 0.012 & 0.012 & 0.000 & 0.003 \\ 
        White Reporting & 0.662 & 0.632 & 0.679 & 0.047 & 0.099*** \\ 
        Estimated Gross Family Income & 107.130 & 105.862 & 107.870 & 2.008 & 0.024 \\ 
        High School GPA & 3.491 & 3.428 & 3.528 & 0.100 & 0.094*** \\ 
        Max ACT Score & 29.831 & 29.575 & 29.980 & 0.406 & 0.124*** \\ 
        Max English \%ile & 62.241 & 59.772 & 63.677 & 3.905 & 0.121*** \\ 
        Max In-House Math Placement & 17.495 & 17.413 & 17.542 & 0.129 & 0.024 \\ 
        Pass FYW on first attempt & 0.990 & 0.989 & 0.991 & 0.002 & 0.024 \\ 
        Grade in FYW & 89.589 & 89.334 & 89.712 & 0.378 & 0.065* \\ 
        Native English Speaker & 1.948 & 1.934 & 1.956 & 0.022 & 0.099*** \\ 
        N & 5619 & 2075 & 3544 & ~ & ~ \\ \hline
    \end{tabular}
    }
    \caption*{\footnotesize \textbf{Notes}: Descriptive statistics for the variables of interest. The effect size columns reports the standardized difference between students below and above the placement recommendation cut-off value. * p $<$ 0.05, ** p $<$ 0.01, *** p $<$ 0.001}
    \label{tbl:placement}
\end{table}



\begin{figure}[h]
  \centering
  \caption{Characteristic variables across the placement threshold}
  
  \begin{minipage}{0.24\textwidth}
    \centering
    \includegraphics[width=\linewidth]{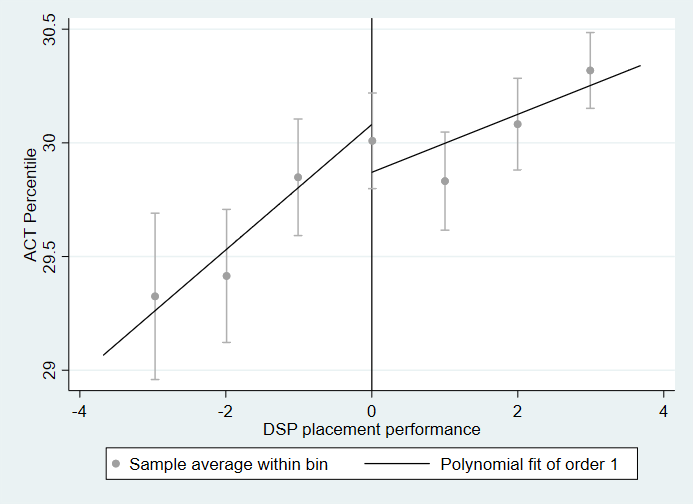}
    \caption*{(a) ACT score}
  \end{minipage}\hfill
  \begin{minipage}{0.24\textwidth}
    \centering
    \includegraphics[width=\linewidth]{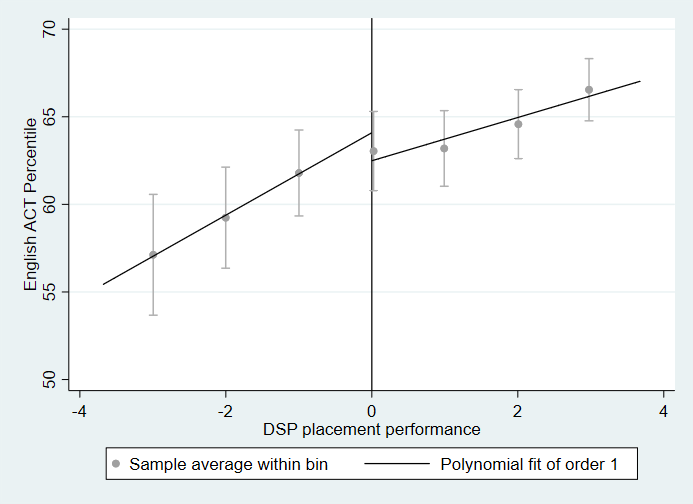}
    \caption*{(b) ACT English}
  \end{minipage}\hfill
  \begin{minipage}{0.24\textwidth}
    \centering
    \includegraphics[width=\linewidth]{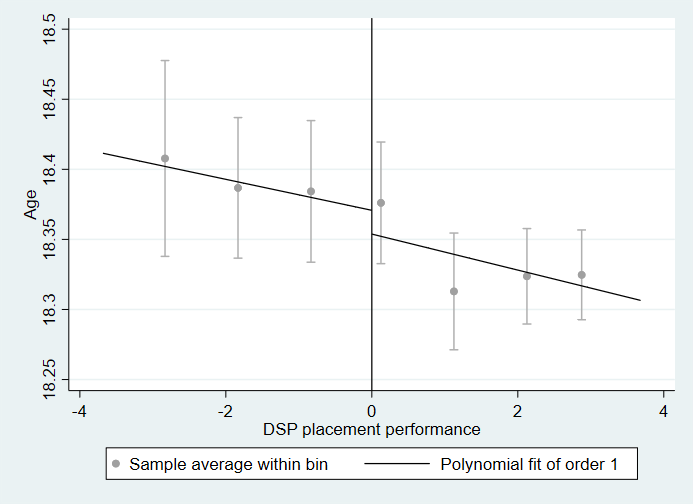}
    \caption*{(c) Age}
  \end{minipage}\hfill
  \begin{minipage}{0.24\textwidth}
    \centering
    \includegraphics[width=\linewidth]{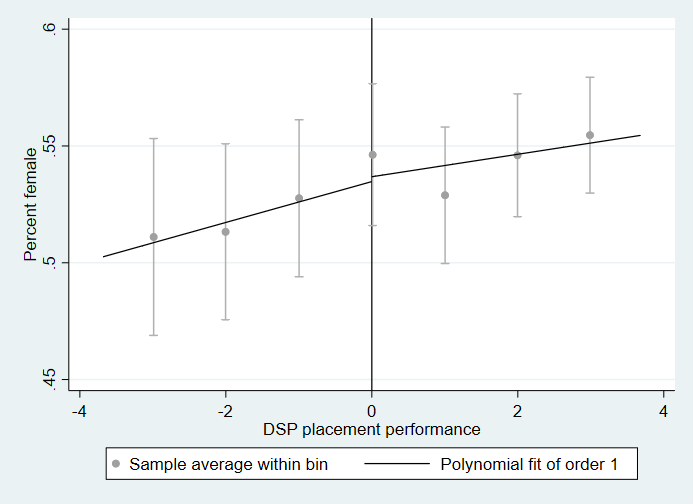}
    \caption*{(d) Female}
  \end{minipage}

  \vspace{0.3cm}
  \begin{minipage}{0.24\textwidth}
    \centering
    \includegraphics[width=\linewidth]{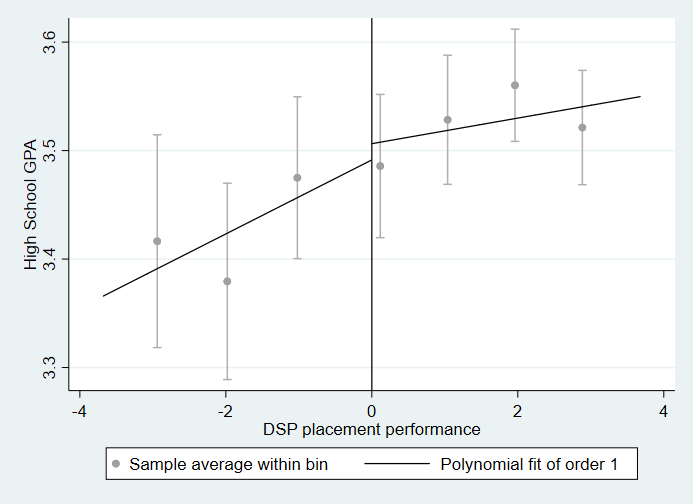}
    \caption*{(e) HS GPA}
  \end{minipage}\hfill
  \begin{minipage}{0.24\textwidth}
    \centering
    \includegraphics[width=\linewidth]{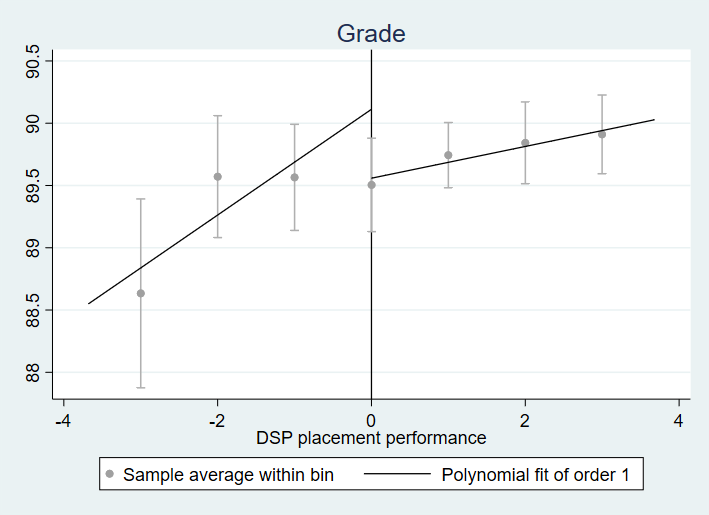}
    \caption*{(f) Grades}
  \end{minipage}\hfill
  \begin{minipage}{0.24\textwidth}
    \centering
    \includegraphics[width=\linewidth]{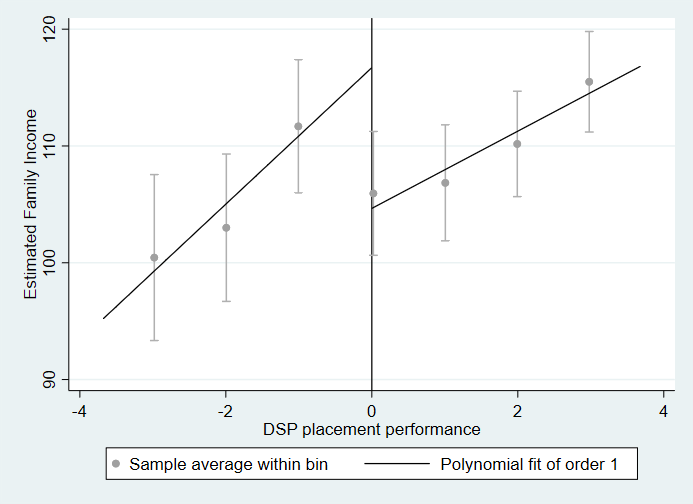}
    \caption*{(g) Family Income}
  \end{minipage}\hfill
  \begin{minipage}{0.24\textwidth}
    \centering
    \includegraphics[width=\linewidth]{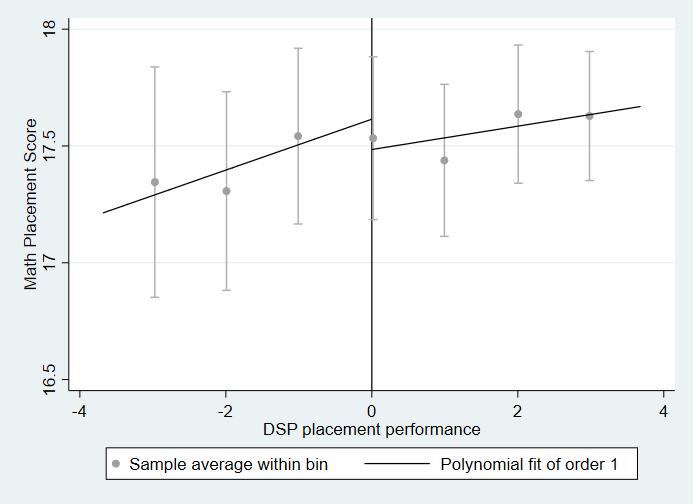}
    \caption*{(h) Math Placement}
  \end{minipage}

  \label{fig:characteristic}
\end{figure}

\begin{figure}[h]
  \centering
  \caption{Outcome variables across the placement threshold}

  \begin{minipage}{0.48\textwidth}
    \centering
    \includegraphics[width=\linewidth]{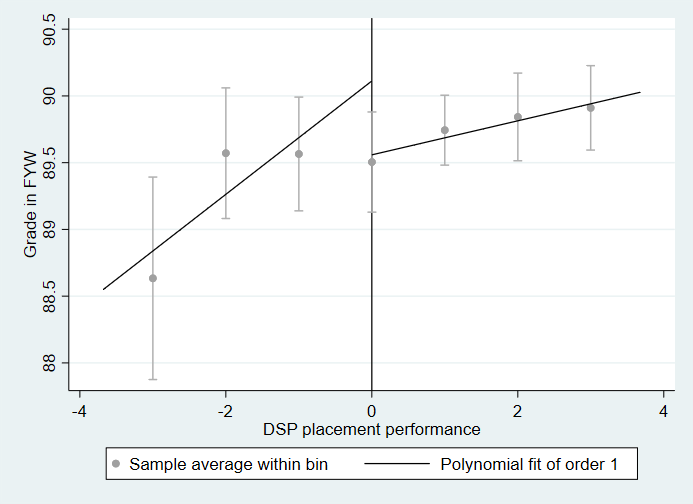}
    \caption*{(a) FYW Grade}
  \end{minipage}\hfill
  \begin{minipage}{0.48\textwidth}
    \centering
    \includegraphics[width=\linewidth]{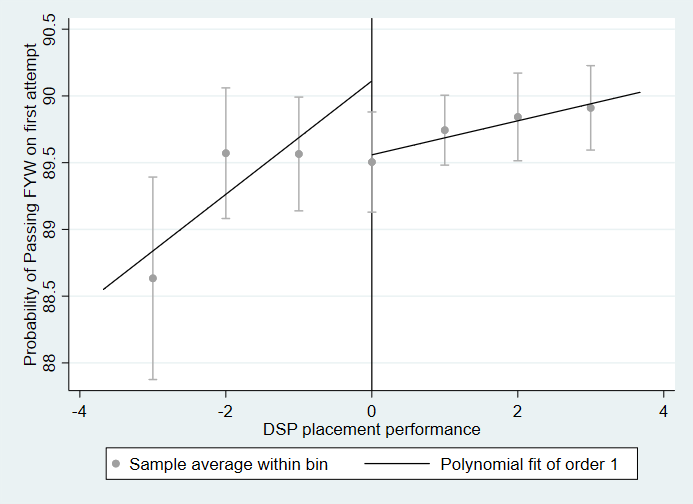}
    \caption*{(b) Probability of Passing}
  \end{minipage}

  \label{fig:outcomes}
\end{figure}

\end{appendices}

\end{document}